\documentclass[twocolumn,showpacs,preprintnumbers,amsmath,amssymb,aps,pof]{revtex4-1}

  \usepackage{graphicx}
  \usepackage{amsmath}
  \usepackage{amsfonts}
  \usepackage{amssymb}
  \usepackage{nicefrac}
  \usepackage{color}

  \newcommand{\vecd}[1]{\mathbf{#1}}
  
  \newcommand{\tens}[1]{\sf {#1}}
  \newcommand{\matr}[1]{\tens{#1}}

\begin{document}

\title{
Wall attraction and repulsion of hydrodynamically interacting particles 
}

\author{Steffen Schreiber$^1$, Jochen Bammert$^1$, Philippe Peyla$^2$ and Walter Zimmermann$^1$}
\affiliation{$^1$ Theoretische Physik I, Universit\"at Bayreuth, 95440 Bayreuth, Germany}
\affiliation{$^2$ Laboratoire Interdisciplinaire de Physique, UMR, Universit\'e Joseph-Fourier, 38402 Saint Martin d'Heres, France}

\date{\today}

\begin{abstract}
We investigate hydrodynamic interaction effects between
colloidal particles in the vicinity of a wall in the low Reynolds-number limit.
Hydrodynamically interacting pairs  of beads being dragged 
by a force parallel to a wall, as for instance during sedimentation, are repelled by the boundary.
If a pair of beads is trapped by harmonic
potentials parallel to the external flow and
at the same distance to a wall, then 
the particle upstream is repelled from the boundary while
its neighbor downstream is attracted. 
 The free end of a  semiflexible bead-spring polymer-model,
which is fixed at one end in a flow near a wall,  is bent towards the wall 
by the same reason.
The results obtained for  point-like particles are exemplarily confirmed by 
fluid particle dynamics simulations of beads of finite radii, where the
 shear induced particle rotations either weaken or enhance 
the effects obtained for point-like particles.

\end{abstract}

\maketitle

\section{Introduction}\label{sec:intro}

The flow properties of suspensions depend very much
on the interaction between particles via the fluid,
the so-called hydrodynamic interaction (HI) 
\cite{Brenner:1991,Russel:1989,Dhont:96,Leal:2007}.
In microfludics, where the distance between
particles and walls becomes often small,  the dynamics of 
 particles  can be strongly influenced  by
the wall-induced  hydrodynamic interaction effects
between rigid
as well as for soft particles, such as
vesicles or polymers. Accordingly, particles
may experience displacements across the unperturbed
streamlines of an external flow and therefore, may lead
to particle redistributions across 
the pipe diameter, if HI and inertia effects are taken into
account. Thus, studies of suspensions are essential
both from the fundamental and from the practical point of view.

In the case of small but finite values
of the Reynolds number,  
Segr\'e and Silberberg
discovered the effect of cross-stream migration
of particles 
to specific positions away from the
centerline of a tube flow
\cite{Silberberg:61.1,Silberberg:62.1,Leal:1980.1,Hinch:1989.1,JosephDD:2005.1,Stone:2009.1}. 
This particle focusing is understood to arise from the
force balance between a wall effect that 
drives the particles to the center of a channel
and a shear-gradient-induced migration pushing
the particles towards the boundary.

In the over-damped Stokes limit in fluid dynamics the interplay between
the HI and the deformability of soft particles like tank-treading
vesicles in shear flows leads to a lift force
close to boundaries \cite{Seifert:99.1,Misbah:99.1,Viallat:2002.1}.
For single soft particles such as oil drops or vesicles, 
their deformability combined with the shear gradient 
in Poiseuille flow causes cross-stream migration
\cite{Leal:1980.1,Kaoui:2008.1}, even in the absence of wall effects.
In these cases the interesting question arises, whether and to which extend
 particle-wall HI affects the cross-streamline migration too \cite{Kaoui:2008.2}. Wall-induced
cross-stream migration may occur also for polymers in suspension in a pipe flow, which is
a long studied and important problem with a number of recent insights 
\cite{Agarwal:1994.1,GrahamM:2004.2,GrahamM:2006.1,GrahamM:2011.1,Chelakkot:2010.1}, or
 during sedimentation \cite{Lundgren:1987} with particle
depletions close to a boundary \cite{Nissila:2004.1} and 
 during active motions of micro-swimmers in confinement \cite{Powers:2009}.
Moreover, it has been found experimentally and theoretically
that swimming microorganisms may be
attracted by solid walls \cite{Lauga:2008}.

Wall-grafted polymer-brushes, which are also used for tuning
surface properties in microchannels, are another example 
 \cite{Minko:2006,Brenner:2005,Stamm:2009}, where the particle-wall interactions
play an important role
for the  dynamics of polymers and which differs significantly from that in the bulk.
 For instance, the    cyclic motion of polymers tethered 
at a surface depends crucially on the interplay between the
 polymer-wall interaction   and the shear flow
 \cite{Viovy:2000,Chu:2005,Delgado:2006.1,dePablo:2009,Shaqfeh:2009,Loewen:2009,AlexanderK:20011.1}.
Here, the HI plays a major role similar as  for the related oscillatory motion of 
three trapped and  hydrodynamically interacting 
particles in shear flow \cite{Holzer:2006.1}.
A related problem is the dynamics of 
cellulose fiber suspensions close to a wall,
which is  important for  paper-manufacturing
and  therefore intensively investigated in order to better control the fiber
orientation \cite{Carlsson_A:2007}.

The examples mentioned so far 
focus mainly on the behavior of single rigid or soft particles (with more
dynamical degrees of freedom) in fluid flow.
For several disconnected particles there are a number of other 
interesting hydrodynamically induced  interaction effects even in the bulk and in 
the absence of fluctuations.
Besides the
oscillatory dynamics of three sedimenting free particles  \cite{Caflisch:1988.1}
and of three trapped particles in shear flow \cite{Holzer:2006.1},
one finds also  HI induced attraction or repulsion between
 asymmetric rotors \cite{Schreiber:2010.1},
and an attraction between tethered polymers in plug flow \cite{Kienle:2011.1}.
For a diluted suspension of Brownian particles in shear flow 
an enhanced self-diffusion
in shear flows is reported \cite{Zarraga:2002.1}, which is
explained by a wall-induced migration of free particles \cite{Wajnryb:2007.1}.

In this work we focus on
boundary induced  hydrodynamic interaction effects between
particles fixed closed to a wall in Stokes flow or dragged particles in a quiescent fluid as described by
the well known technique introduced by Blake for point particles \cite{Blake:1971}
(whereof extensions may be found for instance in 
Refs.~\cite{Wajnryb:2000.1,GrahamM:2004.2,Brady:2007}). The reminder of the
article is organized as follows: 
The related basic equations
of motion for point-like particles are presented in  Sec.~\ref{sec:model},
including a summary of Blake's results. 
 The  results of our numerical and analytical investigations on the
particle-wall HI  
are presented in  Sec.~\ref{sec:results}, where
as a basis of our analysis the flow lines around a single particle
close to a wall are given in Sec.~\ref{sec:single_bead}.
Two or more sedimenting particles near a boundary 
are considered in Sec.~\ref{sec:sedimentation},
where a cross-streamline migration away from the wall
can be qualitatively explained in terms of the flow
lines around a single particle.
The case of two trapped point like particles
exposed to a flow is described in Sec.~\ref{sec:two_beads}.
The results of the two trapped particles 
are qualitatively confirmed by fluid particle dynamics
simulations  \cite{Tanaka:2000,Peyla:2007}  for particles with finite radii
and they illustrate the major effects that are relevant
for the applications to semiflexible bead-spring models 
 in  Sec.~\ref{sec:appli}. In Sec.~\ref{sec:parallel_chain}
a semiflexible bead-spring chain tethered close to a wall and
exposed to a flow is treated and in Sec.~\ref{sec:severalchains}
semiflexible polymers  perpendicularly anchored at a wall in
shear flow.
The article closes in Sec.~\ref{sec:conclusions} 
with a discussion of the results and suggestions of possible experiments.

\section{Model equations}\label{sec:model}

We consider the dynamics of colloidal particles in the limit of a vanishing
Reynolds number, where the laminar flow is described by the
Stokes equation for an incompressible Newtonian fluid.
If not stated otherwise we assume point-like particles
with an effective hydrodynamic radius $a$.

The dynamics of the $N$ beads at 
the positions ${\bf r}_i$ ($i=1, \ldots, N$) is
governed by $N$ coupled equations
\begin{equation}
 \dot{\vecd{r}}_i = {\matr{H}}_{ij} \vecd{F}_j + \vecd{u}_0(\vecd{r}_i) \,,
\label{dynamic_equation}
\end{equation}
where $\vecd{u}_0(\vecd{r}_i)$ is an externally applied flow as for
example a linear shear flow
\begin{equation}
\label{shearflow}
 \vecd{u}_0(\vecd{r})= u_0(z) \hat {\bf x} = \dot \gamma z \hat {\bf x}\, ,
\end{equation}
with the shear rate $\dot \gamma$.
The force $\vecd{F}_j$ is the sum over all potential 
forces acting on the $j$-th bead. Depending on the specific 
system,  these may include stretch or bending forces
as well as forces due to trap potentials,
which can be derived from a potential $V$ according to
\begin{eqnarray}
 \vecd{F}_j & = & - \nabla_j V ,
\label{deterministic_forces}
\end{eqnarray}
where $\nabla_j$ denotes the gradient with respect to $\vecd{r}_j$.
The relevant expressions for $V$ are specified in
Sec.~\ref{sec:results} and Sec.~\ref{sec:appli}.

${\matr{H}}_{ij}$ refers to the mobility matrix describing the
HI between the beads $i$ and $j$.
In the presence of a plane wall with a no-slip boundary condition
for the fluid at the wall, ${\matr{H}}_{ij}$ is given by the Blake tensor \cite{Blake:1971},
\begin{eqnarray}
 {\matr{H}}_{ij}(\vecd{r}_i,\vecd{r}_j) &=&
{}^{S}{\matr{H}}_{ij}(\vecd{r}_i,\vecd{r}_j) -
{}^{S}{\matr{H}}_{ij}(\vecd{r}_i,\vecd{r}_j') \nonumber \\
&+& {}^{D}{\matr{H}}_{ij}(\vecd{r}_i,\vecd{r}_j') -
{}^{SD}{\matr{H}}_{ij}(\vecd{r}_i,\vecd{r}_j') \, ,
\label{mobility_matrices_general}
\end{eqnarray}
where $\vecd{r}'_j = (x_j,y_j,-z_j)$ is the position of
the mirror image of bead $j$ at the opposite side of the 
boundary. The first contribution to ${\matr{H}}_{ij}$ accounts for
the  HI in the unbounded domain described by  the Oseen tensor \cite{Dhont:96},
\begin{equation}
 ^{S}{\matr{H}}^{\alpha \beta}_{ij}(\vecd{r}_i,\vecd{r}_j) = \left\lbrace
\begin{array}{ll}
\frac{1}{8 \pi \eta r_{ij}} \left( \delta_{\alpha \beta} + \frac{r_{ij}^{\alpha}  r_{ij}^{\beta}}{r_{ij}^2} \right) & \quad \mathrm{for} \, i \neq j  \,, \\
\frac{1}{6 \pi \eta a} \, \delta_{\alpha \beta} & \quad \mathrm{for} \, i = j  \,,
\end{array}
\right.
\label{H_oseen_1}
\end{equation}
and the second one for the HI between the 
beads and their mirror images,
\begin{equation}
 ^{S}{\matr{H}}^{\alpha \beta}_{ij}(\vecd{r}_i,\vecd{r}_j') = \frac{1}{8 \pi
\eta \tilde{r}_{ij}} \left( \delta_{\alpha \beta} +
\frac{\tilde{r}_{ij}^{\alpha} \tilde{r}_{ij}^{\beta}}{\tilde{r}_{ij}^2} \right) \,,
\label{H_oseen_2}
\end{equation}
where $\eta$ is the viscosity of the fluid.
We furthermore use the abbreviations
\begin{subequations}
\begin{eqnarray}
 \vecd{r}_{ij} &=& \vecd{r}_i - \vecd{r}_j = r_{ij} \hat{\vecd{r}}_{ij} \,,\\
 \tilde{\vecd{r}}_{ij} &=& \vecd{r}_i - \vecd{r}'_j = \tilde r_{ij} \hat{\tilde{\vecd{r}}}_{ij} \,\;
\end{eqnarray}
\end{subequations}
and the components of the vector $\vecd{r}_{ij}$  ($\tilde{\vecd{r}}_{ij}$)
are denoted by $r_{ij}^\alpha$ ($\tilde r_{ij}^\alpha$), where $\alpha=x,y,z$.
In Eq.~(\ref{mobility_matrices_general}) the contribution
\begin{equation}
^{D}{\matr{H}}_{ij}^{\alpha \beta}(\vecd{r}_i,\vecd{r}_j') = \frac{1}{4
\pi \eta \tilde r_{ij}^3} \, z_j^2 (1-2\delta_{\beta z}) \left( \delta_{\alpha \beta} -
3 \frac{\tilde r_{ij}^{\alpha} \tilde r_{ij}^{\beta}}{\tilde r_{ij}^2} \right)
\label{H_oseen_3} 
\end{equation}
is the Stokes doublet ($D$) and
\begin{eqnarray}
^{SD}{\matr{H}}_{ij}^{\alpha \beta}(\vecd{r}_i,\vecd{r}_j') & = & \frac{1}{4
\pi \eta \tilde r_{ij}^3} \, z_j (1-2\delta_{\beta z}) \nonumber \\
&& \hspace{-1,8cm} \left( \delta_{\alpha \beta}
\tilde r_{ij}^z - \delta_{\alpha z} \tilde r_{ij}^{\beta}   + \delta_{\beta z} \tilde r_{ij}^{\alpha} -
3 \frac{\tilde r_{ij}^{\alpha} \tilde r_{ij}^{\beta} \tilde r_{ij}^z}{\tilde r_{ij}^2} \right)
\label{H_oseen_4}
\end{eqnarray}
is the source doublet ($SD$). In our numerical calculations higher order
corrections to ${\matr{H}}_{ij}$ due to the finite size of the
spheres are included up to the order $a^2$ (Rotne-Prager approximation).
The final equations of motion are given by
\begin{equation}
 \dot{\vecd{r}}_i = \tilde{ \matr{H}}_{ij} \vecd{F}_j +
\vecd{u}_0(\vecd{r}_i) \,,
\label{method_of_reflections}
\end{equation}
where  the  mobility matrices 
\begin{equation}
\label{mobmat}
\tilde{\matr{H}}_{ij}=\left( 1+\frac{a^2}{6} \nabla_i^2 +\frac{a^2}{6} \nabla_j^2 \right) {\matr{H}}_{ij}
\end{equation}
 fulfill the relation $\matr{H}_{ij} = \matr{H}_{ji}^{T}$,
which is important for the overall symmetry of the problem -
as pointed out in Ref.~\cite{Brady:2007}.

\section{Results for basic models}\label{sec:results}

In this section we determine the influence of
wall-induced hydrodynamic interactions on the dynamics of
two and three beads which are either dragged in a fluid
parallel to a boundary or hold in potentials near a
wall and exposed to flows.
To this end Eq.~(\ref{method_of_reflections}) 
with Eq.~(\ref{mobmat})
is solved numerically for the  viscosity $\eta=1$ 
and approximate analytical
solutions are given in some cases.

%
\subsection{A single trapped bead in shear flow close to a wall}\label{sec:single_bead}
%

The influence of a wall on the HI
between two beads can be illustrated by the
streamlines around one point-like particle with an 
effective hydrodynamic radius $a$ at
a distance $d_w$ from a solid boundary.
The particle is exposed to the flow given by Eq.~(\ref{shearflow})  and is trapped 
by a harmonic potential
\begin{equation}
 V = \frac{k_{\mathrm{pot}}}{2} (\vecd{r}_b-\vecd{r}_{\mathrm{pot}})^2 \, ,
 \label{potential_single_bead}
\end{equation}
where $\vecd{r}_b$ is the position vector of the particle,
$\vecd{r}_{\mathrm{pot}}$ the position of the
potential minimum, with a spring constant of
 $k_{\mathrm{pot}} = 1$.
Equivalently, one could consider a point-like particle dragged by
a constant external force ${\bf F}=f\hat {\bf x}$ parallel to the wall
in a quiescent fluid.
The streamlines around the point-like particle are the same 
in both cases if a comoving frame is chosen,
where the particle's position is held fixed by the potential at
 $\mathbf{r}_{pot}=(0,0,d_w)$.

The trap force ${\vecd F}^s = -k_{\mathrm{pot}}\left(\vecd{r}_b-\vecd{r}_{\mathrm{pot}}\right)$
as a function of the unperturbed flow velocity ${\bf u}_0(\bf{r})$ 
is obtained by solving Eq.~(\ref{method_of_reflections}) for
a vanishing bead velocity $\dot {\vecd r}_b=0$: 
\begin{equation}
 0 =  \tilde{{\matr{H}}} (\vecd{r}_b, \vecd{r}_b) \vecd{F}^s + \vecd{u}_0(\vecd{r}_b) \,.
\label{force_fixed_bead}
\end{equation}
We find for the Oseen approximation  
\begin{equation}
 F^s_x = f_x~\frac{16 d_w}{16d_w -9a} \qquad \mathrm{and} \qquad  F^s_y=F^s_z=0 \,,
\label{potential_force}
\end{equation}
 with $f_x=-6\pi \eta a u_0(d_w)$  the force required to keep the
particle fixed in the absence of a wall, if it is exposed to a flow
with the velocity $u_0(d_w)$.
When $u_0(d_w)$ is constant 
for varying $d_w$, {\it e.g.}, by adjusting
the shear rate appropriately,
the force $F^s_x$ exerted on the bead
increases with decreasing $d_w$. This effect
will be investigated further in subsection \ref{sec:two_beads}.

The vanishing force $F^s_z=0$ perpendicular to the wall
in Eq.~(\ref{potential_force}) 
reflects the time reversibility of the Stokes equation \cite{Brenner:1991}.
If a single bead in shear flow would migrate
perpendicularly to the wall, {\it i.e.} $F_z^s \neq 0$,
then for symmetry reasons the drift would point  
in the same direction after reversing the flow.
But then the
motion would not be reciprocal and the time-reversibility of the 
Stokes equation would be violated. Therefore, $F^s_z \not =0$ is forbidden 
and there is no migration of a single bead perpendicular to the wall.

\begin{figure}[ht]

     \includegraphics[width=0.72\columnwidth, angle = -90]{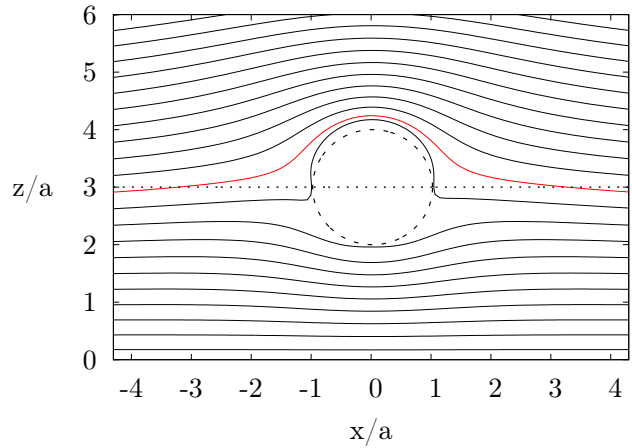}
  \caption{In the $xz$ plane the streamlines around
the bead fixed at ${\vecd r}_b=(0,0,d_w=3)$  
are in the presence of a plane no-slip boundary at $z=0$
 asymmetric with respect to the axis at $z=d_w$. 
One obtains the same asymmetry of the flow lines in the co-moving frame
of a point particle dragged parallel to the boundary along the dotted line at $z=d_w$.
The flow lines were calculated via Eq.~(\ref{flow_around_bead}),
whereby the red line marks the trajectory of a test particle starting and terminating
below the center of the bead but passing the bead on top.
}
\label{streamlines}
\end{figure}

The flow field ${\vecd u}({\vecd r}) $ around a bead fixed at 
${\vecd r}_b = (0,0,d_w)$  shown in Fig.~\ref{streamlines}
 is given by 
\begin{equation}
{\vecd u}({\vecd r})   =  \left(1+\frac{a^2}{6} \Delta \right)  {\matr{H}} ( \vecd{r}, \vecd{r}_b) \vecd{F}^s +
\vecd{u}_0(\vecd{r}) \,.
\label{flow_around_bead}
\end{equation}
 In the absence of a boundary, where only the 
first contribution in Eq.~(\ref{mobility_matrices_general}) 
has to be taken into account, the streamlines are up/down and left/right
symmetric with respect to the center of the bead. 
In the presence of a wall this up/down 
symmetry is broken and the streamlines are deformed 
as indicated in Fig.~\ref{streamlines}. This effect is caused by the HI 
between the fixed bead and the boundary, which is described 
by the second, third, and fourth contribution to 
the mobility matrix in Eq.~(\ref{mobility_matrices_general}).
If one introduces a small tracer particle, starting 
at $x<0$ at a height $z<d_w$, it follows one of the displayed
 streamlines as indicated by the
red streamline in Fig.~\ref{streamlines}
 and may pass the bead at the side opposite from the wall, {\it i.e.} $z>d_w$.
According to the $\pm x$ symmetry the streamline reaches the range $z<d_w$ again
for large $x$. This is a consequence of  
the time-reversibility of the Stokes equation, which is also valid in the presence of a solid boundary.
The  wall-induced deformation of the streamlines 
has interesting consequences as discussed in the following sections.

\subsection{Particles dragged parallel to a boundary - a model for sedimentation}\label{sec:sedimentation}

Here we investigate the motion of several particles which are
dragged by a constant external force ${\bf F}$
parallel to a wall, as for example by the  gravitational force
during the sedimentation of particles.

\begin{figure}[ht]
  \begin{center}
    \includegraphics[width=\columnwidth]{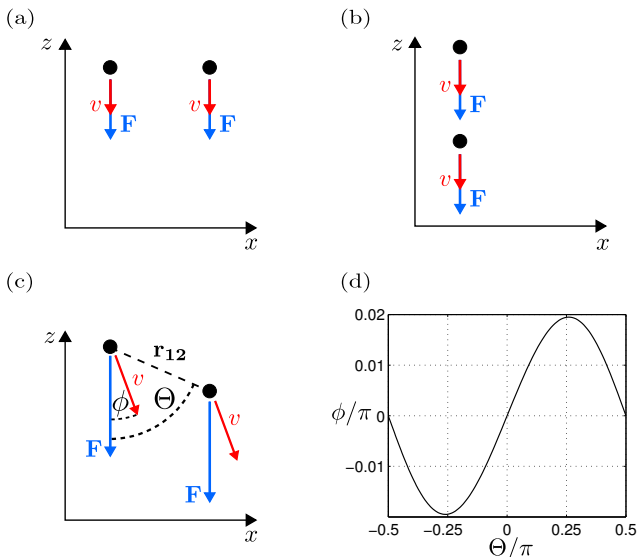}
  \end{center}
  \caption{Two particles are dragged through a fluid
by a force ${\bf F}$ anti-parallel to the vertical $z$ axis. 
The beads move parallel to ${\bf F}$, if the 
connection vector $\vecd{r}_{12}$ encloses with ${\bf F}$ an angle 
$\Theta = \pi/2$, as in part  (a), or $\Theta = 0$, as in  part (b). 
For other values of $\Theta$ the HI effects cause an angle $\phi$ between the direction of the
particle velocity ${\bf v}$ and $\bf{F}$, as indicated in  (c) \cite{Brenner:66}. An
approximation of the dependence of $\phi$ on $\Theta$
is given by Eq.~(\ref{phi_theta_bulk}) and plotted in (d).
}
\label{bulkmov}
\end{figure}

We first consider the motion of two beads dragged in
an unbounded fluid. If the force ${\bf F}$ acts parallel or perpendicular to the
connection vector between both particles, $\vecd{r}_{12} = \vecd{r}_1 - \vecd{r}_2$,
cf. Fig.~\ref{bulkmov}(a) and (b),
they move in either case parallel to the force,
as indicated in Fig.~\ref{bulkmov} by part a) and b).
If $\vecd{r}_{12}$ encloses an angle $\Theta \not =0$ and
$\Theta \not=\pi/2$ with the force ${\bf F}$, the particles move obliquely
with respect to the drag force due to
the HI between the beads as indicated in 
 Fig.~\ref{bulkmov}(c) (see also \cite{Brenner:66}).
The deflection angle $\phi$ between the direction of the particle motion and the force
has its maximum value at $\Theta = \pm \pi/4$ and is in the Oseen approximation given by
\begin{equation}
 \phi = \arctan{ \left( \frac{\sin{\Theta} \cos{\Theta}}{1+\frac{4d}{3a}+\cos^2{\Theta}} \right) } \,,
 \label{phi_theta_bulk}
\end{equation}
where $d = |\vecd{r}_{12}|$ is the distance between the particles.

\begin{figure}[ht]
  \begin{center}
    \includegraphics[width=0.95\columnwidth]{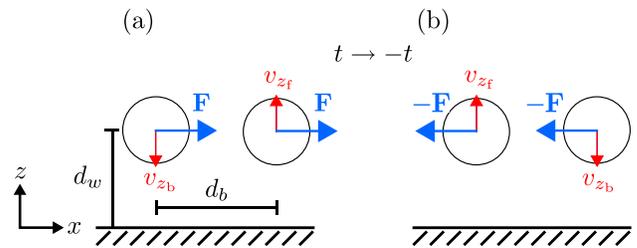}
  \end{center}
  \caption{Each of the two beads in part a)
is dragged by a force ${\bf F}$ parallel to a wall.
The particle in front is repelled from the wall, 
while the bead behind is attracted, as expected by the form of the
 streamlines around a single bead in
Fig.~\ref{streamlines}. In part b) the time has been reversed 
and thus the direction of the applied force and the particle motion
are reversed too.}
\label{symmetry_consideration_2beads}
\end{figure}

\begin{figure}[ht]
  \begin{center}
    \includegraphics[width=0.7\columnwidth,angle=-90]{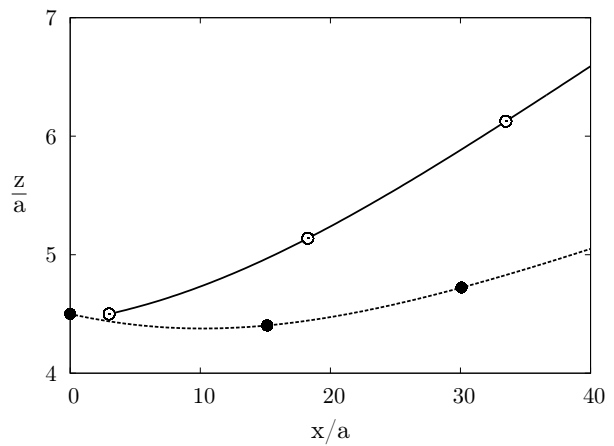}
  \end{center}
  \caption{
The trajectories of two beads in the $xz$ plane are shown, 
which are dragged through the fluid by an external force ${\bf F} \parallel \hat {\bf x}$
parallel to the boundary at $z=0$. 
The initial positions of both beads are 
located at ${\bf r}_1=(0,0,d_w)$
and   ${\bf r}_2=(d_b,0,d_w)$ with $d_b=3a$ and $d_w=4.5a$.
The particle positions are indicated by open and filled 
circles along the trajectories at equal times. 
The bead behind (dashed line) is at first attracted towards the wall
due to the wall-mediated HI,
whereas the particle in front (solid line) is always repelled. Later on
the bulk HI dominates and both particles move away from the wall, as expected
according to Eq.~(\ref{phi_theta_bulk}).
}
\label{sedimentation_2_beads_parallel}
\end{figure}

In the presence of a boundary   
the particle-particle hydrodynamic interaction  via the boundary comes  into play.
As already indicated by the flow lines around a single particle in  
Fig.~\ref{streamlines}, a wall breaks the symmetry around the
particle with respect to the $z$ direction.
For two particles, which are initially located at the same distance from a wall
and pulled parallel to it, as depicted in Fig.~\ref{symmetry_consideration_2beads}(a),
we show in Fig.~\ref{sedimentation_2_beads_parallel} the trajectories 
of point particle's determined numerically from Eq.~(\ref{method_of_reflections}).
The open and the filled circles in Fig.~\ref{sedimentation_2_beads_parallel} represent the
particle positions at equal times.

The bead in front is repelled by the wall immediately from the start, while the
particle behind is at first attracted to the wall,
as indicated in Fig.~\ref{symmetry_consideration_2beads}a). Later on
the rear bead  moves away from the wall as well.
In order to understand this effect it is useful to consider the early and
the later stage of the motion shown in Fig.~\ref{sedimentation_2_beads_parallel} 
for each particle.
For the early regime the flow-lines around a single bead, as shown in Fig.~\ref{streamlines},
allow an estimate about the motion of the two particles 
along the $z$ direction. In the front of a pulled particle the velocity of the fluid has 
a component in the positive $z$ direction and in the negative $z$ direction behind it.
Accordingly, in the case of a pair of dragged beads the particle in front
is repelled from the wall
whereas the one behind is attracted towards 
the boundary as sketched in Fig.~\ref{symmetry_consideration_2beads}(a).
During this process of wall-particle repulsion and attraction  
the connection vector $\vecd{r}_{12}$ becomes skewly oriented to
the drag force, i. e. $\Theta \neq 0$ as indicated in Fig.~\ref{bulkmov}c), with the
particle closer to the wall behind the other.

If the connection vector is skewly oriented,  the bulk effect, as described above, comes into play in
the second stage of the motion. It
causes in the case, when the particle closer to the wall
moves behind the other one, a drift of both particles perpendicular
to the external force and away from the wall.
The  bead closer to the wall moves slower than the other one because its
effective friction is enhanced closer to the wall  (cf. Eq.~(\ref{potential_force}))
and therefore the bead distance $d$ increases and $\Theta$ decreases.
Consequently $|\phi|$ decreases in agreement with Eq.(\ref{phi_theta_bulk}).
In the long-time limit the particles'  trajectories align with the applied forces,
in which case the wall distance of both particles practically saturate.
Therefore, the essential effect of the
hydrodynamic particle-wall-particle interaction on the motion of the beads
 shown in Fig.~\ref{sedimentation_2_beads_parallel}
is the reorientation of their connection vector $\vecd{r}_{12}$.
When $\vecd{r}_{12}$ is
oblique  to the drag force, the bulk effect provides the major contribution 
to the migration of the particles away from the wall.

Near a  point-like particle at $\vecd{r} = (0,0,d_w)$, which is
 dragged parallel to a  boundary, the induced velocity of the
fluid around it can be determined analytically  by taking  
the wall effects into account in the Oseen-approximation.
In this case  the non-zero $z$-component of the velocity of the
fluid at the positions $\vecd{r}_{\pm} = (\pm d_b,0,d_w)$
is as follows:
\begin{equation}
  v_{z_\pm}=\pm \frac{3}{2} \frac{f}{\pi \eta} \frac{d_b d_w^3}{(4d_w^2+d_b^2)^{5/2}} \,.
\label{2_beads_initial_velocity}
\end{equation}
If a free test
particle is placed at $\vecd{r}_{\pm}$ near the dragged
particle, it moves with the fluid and its induced vertical
velocity component is given by the 
fluid velocity $v_{z_{\pm}}$. These
velocity components have the same magnitude in the front
and in the rear of the dragged particle, whereas they
point in opposite directions as indicated also by 
the streamlines shown in Fig.~\ref{streamlines}. 
According to this qualitative
reasoning one expects in the case of a  pair of beads, which are 
dragged parallel to a wall, that the particle in front is repelled from
the wall whereas the rear one is attracted towards the wall.

Eq.~(\ref{2_beads_initial_velocity})  displays also the
reciprocity of trajectories in Stokes flow.
If the time and therefore the direction of the applied forces
are reversed, then in the discussed setup the bead in front is again repelled from the wall 
and the rear one is driven to the wall,
as indicated in Fig.~\ref{symmetry_consideration_2beads}(b).
After the time-reversal the role of the beads is interchanged, with 
$v_{z_+} = - v_{z_-}$ and the motion  is reciprocal to the one before, thus obeying
the time reversibility of the Stokes-equation.

\begin{figure}[ht]
  \begin{center}
    \includegraphics[width=0.7\columnwidth,angle=-90]{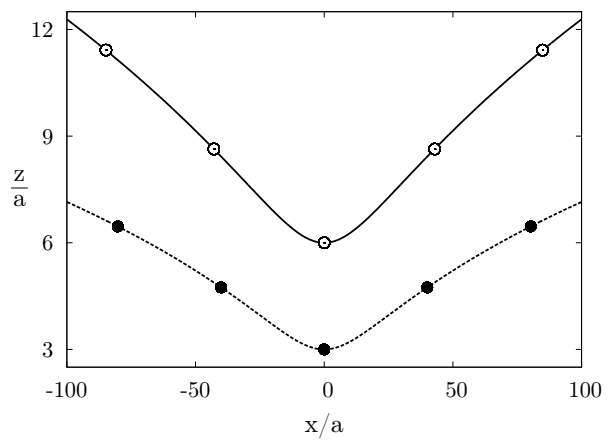}
  \end{center}
  \caption{The trajectories of two beads in the $xz$ plane,
which are dragged by an external force ${\bf F} \parallel \hat {\bf x}$ parallel to
the boundary at the $xy$ plane, {\it i.e.} $z=0$. 
The initial bead connection vector $\vecd{r}_{12}$ is slightly tilted with respect to the $z$ axis
so that the bead which is closer to the wall is slightly in front of the other one.
The open and the filled circles visualize the positions of the two particles
at equal times. First the beads approach the wall until the boundary effect causes a
reorientation of $\vecd{r}_{12}$, so that the particles are finally repelled from
the wall.}
\label{sedimentation_2_beads_perpendicular}
\end{figure}

Fig.~\ref{sedimentation_2_beads_perpendicular} shows 
the trajectories of two particles as obtained by 
integrating Eq.~(\ref{method_of_reflections}) for two particles  dragged 
by a force $\vecd{F} \parallel \hat {\bf x}$ along a wall at $z=0$
and interacting via the Rotne-Prager approximation.
The bead connection vector $\vecd{r}_{12}$ at the initial position
encloses an angle $\Theta < \pi/2$ with $\vecd{F}$.
In this case, the bead  closer to the wall, {\it i.e.} $z=0$, is slightly in front of the other one,
and due to the bulk effect described above (cf. Fig.~\ref{bulkmov}), both beads first approach the wall.
As the HI with the boundary and thus the friction becomes stronger the 
 bead closer to the wall becomes slower than the more distant particle 
(cf. Eq.~(\ref{potential_force})). Consequently, the particle further from the boundary
finally overtakes the other one and in this case
 $\vecd{r}_{12}$ becomes again oblique with respect to
 $\vecd{F}$ with $\Theta > \pi/2$ leading to a repulsion
of both beads from the wall due to the bulk effect.

 Besides the examples in  Fig.~\ref{sedimentation_2_beads_parallel}
and Fig.~\ref{sedimentation_2_beads_perpendicular} we investigated  many other orientation angles  between
$\vecd{r}_{12}$ and $\vecd{F}$ ($\Theta=0$, $\Theta=\pi/2$, $\Theta \gtrless \pi/2$). 
 We obtain for all orientations of the connection vector finally a  
 drift of both particles away from the wall.
Hence, the wall-induced hydrodynamic particle-wall-particle 
interaction between two beads being dragged by
a force parallel to a boundary eventually causes a repulsion
of the particles from the wall for arbitrary initial conditions.

\begin{figure}[ht]
  \begin{center}
    \includegraphics[width=0.7\columnwidth,angle=-90]{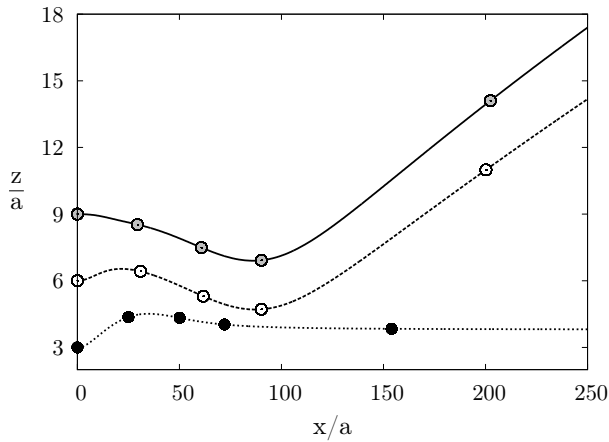}
  \end{center}
  \caption{
The trajectories of three beads in the $xz$ plane,
which are dragged by an external force ${\bf F} \parallel \hat {\bf x}$
away from their initial positions: $(0,0,3a)$, $(0,0,6a)$ and
$(0,0,9a)$.
The positions of the particles at equal times are indicated by open, grey and black filled 
circles along the trajectories. 
Due to the interplay of bulk and wall HI effects 
the upper beads form a pair after an intermediate regime
and finally drift away from the wall due to bulk HI effects.
The third bead remains nearly at its initial distance from the wall.
}
\label{sedimentation_3_beads_perpendicular}
\end{figure}

The trajectories of three particles which are initially
aligned perpendicularly to a wall are shown in Fig.~\ref{sedimentation_3_beads_perpendicular}.
The initial distances between the beads and between the lowest particle and the wall are $3a$.
As explained before, the bead which is furthest from the boundary is often slightly
faster than the other two.
Due to the bulk HI between the particles,
which then have different $x$ positions, the lowest and the highest bead
perturb the flow in such a way that the middle particle experiences 
friction just as small as for the upper one. For the example in 
Fig.~\ref{sedimentation_3_beads_perpendicular} the upper two particles
finally build a pair and move away from the wall in a similar manner as in the case
of two beads described above 
in Fig.~\ref{sedimentation_2_beads_perpendicular}, whereas the
lower (third) particle moves similar to a single particle nearly parallel
to the wall. For other initial conditions similar formations of pairs of beads
are found, which move finally away
from the wall.


The motion of an assembly of many particles is governed by the same principles as described above.
By a complex interplay of bulk and wall effects pairs of beads form and dissolve, but the overall
tendency is a migration of the particles away from the wall, similar to
the results reported in Ref.~\cite{Nissila:2004.1}, where the situation of
small but finite values of the Reynolds number is investigated.

Another case, where the hydrodynamic particle-wall-particle
interaction has to be considered,
is when two beads are dragged perpendicularly to a wall
with identical wall distances. This is equivalent to the situation of
two sedimenting particles, when they approach the bottom boundary.
In this case, the two beads repel each other slightly and
the repulsion becomes significant at wall distances smaller than $10a$.
Possible experimental setups for investigations of the effects
predicted in this subsection are discussed in Sec.~\ref{sec:conclusions}.

\subsection{Two trapped beads in shear flow close to a wall}\label{sec:two_beads}

In this section wall-induced HI
effects are investigated for a system composed 
of two trapped particles in a linear shear flow
near a wall.
The harmonic trap potential is given by 
\begin{equation}
V = \frac{k_{\mathrm{pot}}}{2} \left[ (\vecd{r}_1 - \vecd{r}_{\mathrm{pot},1} )^2 + (\vecd{r}_2 - \vecd{r}_{\mathrm{pot},2})^2 \right] \,,
 \label{potential_two_beads}
\end{equation}
where $k_{\mathrm{pot}} = 1$ is the spring constant and
$\vecd{r}_1$ and $\vecd{r}_2$  are the positions of the two beads. The
locations of the potential minima are
$\vecd{r}_{\mathrm{pot},1} = (0,0,d_w)$ and $\vecd{r}_{\mathrm{pot},2} = (d_b,0,d_w)$
with the connection vector 
$\vecd{r}_{\mathrm{pot},1}- \vecd{r}_{\mathrm{pot},2}$ parallel to the wall.
The results discussed are obtained for point-like particles and compared
to the case of finite-sized beads which undergo in addition shear-induced
rotation.
\begin{figure}[ht]
  \begin{center}
    \includegraphics[width=0.7\columnwidth]{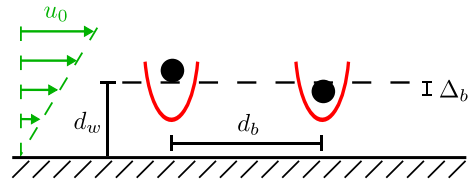}
  \end{center}
  \caption{Two point-like particles are trapped in a shear flow ${\mathbf{u_0}}(z)$
 by harmonic potentials with their  minima at a distance $d_w$ from the wall at $z=0$
 and a mutual separation $d_b$. With the coordinate $z_2$
of the particle downstream  $\Delta_b=d_w-z_2$ measures its 
displacement  from the potential minimum.
}
\label{2_beads_sketches}
\end{figure}

The two point-like particles
are displaced by the flow in $x$ direction away
from the minima of the trapping potential. Simultaneously, 
the wall-induced HI causes displacements along the 
$z$ direction until the influence of the boundary is balanced by the potential forces. 
Similar to the previous section, the bead upstream is effectively repelled from the wall 
and the bead downstream is attracted.
Note, that the deflection of the particle downstream
is different from the displacement of the bead upstream,
which is due to the bulk effect described in the previous
 subsection \ref{sec:sedimentation} and can be explained as follows:
Because of the finite angle $\Theta$ between the bead connection vector
and the trap forces, cf. Fig.~(\ref{bulkmov}),
both particles are shifted away from the wall. 
The boundary effect dominates, because the trap forces prevent the
angle $\Theta$ from becoming too large,
but due to the described shift the bead downstream is attracted towards the wall by a smaller
distance than the one upstream is repelled from the wall.

The wall-induced particle 
displacement along the $z$ direction is characterized by the shift
\begin{eqnarray}
 \Delta_b = d_w - z_2\,,
 \label{defdelta_b}
\end{eqnarray}
where $z_2$ is the steady-state position of the bead downstream.

In the following we determine the displacement $\Delta_b$
at the steady-state as a function of the flow velocity as well as of the
distances $d_w$ and $d_b$.
 As a first approach, $\Delta_b$ is approximated analytically 
for two point-like particles as shown in Fig.~\ref{2_beads_sketches}. 
The forces needed to  hold  the beads in place in an external flow can be
estimated in a similar manner as described in Sec.~\ref{sec:single_bead}, cf. Eq.~(\ref{potential_force}).
In the Oseen approximation 
their $x$ components are given by
\begin{equation}
F^s_x = f_x \frac{16 d_b d_w}{8d_w(2 d_b +3a) - 9 a d_b} \,
 \label{force_two_fixed_beads}
\end{equation}
with $f_x = -6 \pi \eta a u_0(d_w)$. 
For the particle downstream 
one obtains the initial velocity $v_{z_-}$
in $z$ direction via Eq.~(\ref{2_beads_initial_velocity}) with $f = F^s_x$.
The resulting Stokes force 
$F^s_z = 6 \pi \eta a v_{z_-}$ can be used to
estimate the elongation $\Delta_b$ in the steady state via the 
counteracting force $F^{\mathrm{pot}}_{z} = k_{\mathrm{pot}} \Delta_b$,
which must fulfill $F^s_z + F^{\mathrm{pot}}_{z} = 0$. This calculation yields
\begin{equation}
\Delta_b = \frac{864 \pi \eta a^2 d_b^2 d_w^4 u_0(d_w)}{k_{\mathrm{pot}} (4 d_w^2 + d_b^2)^{5/2} ~ \left[ 8d_w( 2 d_b +3a) - 9a d_b  \right]} \,.
 \label{delta_analytical}
\end{equation}
$\Delta_b$ depends linearly on the unperturbed flow
velocity $u_0(d_w)$, which itself increases linearly as a function of $d_w$
according to Eq.~(\ref{shearflow}).
Since we are mainly interested in the $d_w$-dependence of the
wall-induced  HI effects and not in its dependence on the
absolute value of $u_0(d_w)$,
we adjust the shear rate 
according to the relation $\dot \gamma = u_0(d_w)/d_w$, such
that the flow velocity $u_0(d_w)$ becomes independent of $d_w$:
\begin{equation}
\vecd{u}_0 = u_0(d_w) \frac{z}{d_w} \; \hat{\vecd{x}}\,.
\label{shearflow_dw}
\end{equation}
\begin{figure}[ht]
  \begin{center}
    \includegraphics[width=0.7\columnwidth, angle = -90]{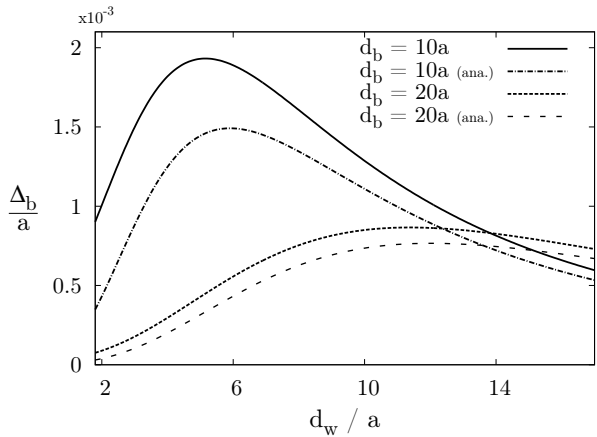}
  \end{center}
  \caption{The bead deflection $\Delta_b$
is plotted as a function of the wall
distance $d_w$ for two different distances $d_b=10a,~20a$ between the potential minima,
 as indicated in Fig.~\ref{2_beads_sketches},
and for the flow field given 
by Eq.~(\ref{shearflow_dw}) with $u_0(d_w)=0.004$.
The dash-dotted line (second from above) and the dashed line
(lowest one) correspond to the
analytical approximations according to Eq.~(\ref{delta_analytical}).
}
\label{delta_dwall_2beads}
\end{figure}

The analytical expression for $\Delta_b$ in Eq.~(\ref{delta_analytical})
exhibits characteristic maxima as a function of $d_w$ and as a function of $d_b$.
The dependence of $\Delta_b$ on $d_w$ is shown in 
Fig.~\ref{delta_dwall_2beads} for the two different 
values  $d_b=10a, ~20a$. In this figure 
the analytical results are compared with the numerically determined values,
which were obtained via Eq.~(\ref{method_of_reflections}).
Despite the assumptions included in Eq.~(\ref{delta_analytical}),
both curves agree with each other surprisingly well qualitatively, 
especially at larger distances $d_w$ and $d_b$.

The maximum of $\Delta_b(d_w)$ in Fig.~\ref{delta_dwall_2beads} can be traced back
to Eq.~(\ref{2_beads_initial_velocity}), which represents the $z$ component of the
perturbed fluid velocity due to a dragged particle near a wall.
This expression already exhibits a maximum as a function of $d_w$.
However, for very large wall distances $d_w \gg d_b$ the influence of the
boundary must vanish and $\Delta_b$ must approach $0$.
\begin{figure}[ht]
  \begin{center}
    \includegraphics[width=0.7\columnwidth, angle = -90]{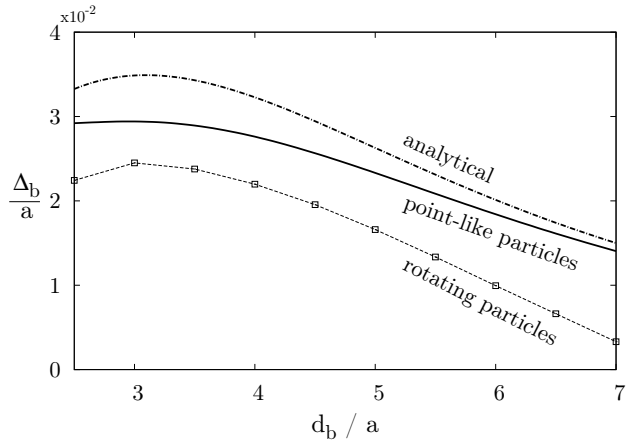}
  \end{center}
  \caption{$\Delta_b$ is plotted as a function of the bead distance $d_b$
for the setup shown in Fig.~\ref{2_beads_sketches} and Fig.~\ref{sketch_fpd}.
The distance to the wall is  $d_w = 2.5a$ and the flow velocity is
$u_0(d_w) = 0.004$. The solid line represents the results for point-like particles,
where the effects of particle rotation are neglected.  
The squares are the result of simulations for particles of a finite size,
which rotate in the external shear flow (cf. Fig.~\ref{sketch_fpd}).
The dash-dotted line is the result of the analytical approximation given by Eq.~(\ref{delta_analytical}).
}
\label{delta_dbead_2beads}
\end{figure}
The dependence of $\Delta_b$ on the particle-particle   
distance  $d_b$ is shown in Fig.~\ref{delta_dbead_2beads}
for $d_w=2.5a$ and for the flow velocity $u_0(d_w) = 0.004$.
The maximum of $\Delta_b(d_b)$ in
the numerical solution (solid line),
is less pronounced than in the corresponding
analytical expression (dash-dotted line).

\begin{figure}[ht]
  \begin{center}
    \includegraphics[width=0.75\columnwidth]{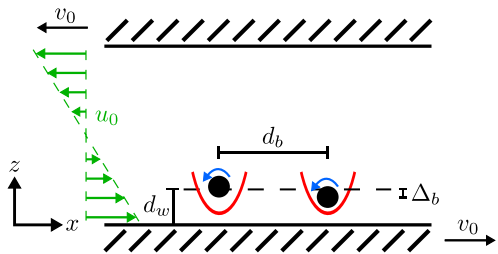}
  \end{center}
  \caption{The shear-induced 
particle rotation influences the
wall-induced particle attraction and repulsion.
For the given setup the deflection of both beads from their initial position
$z=d_w$ is reduced due to the rotation effects.
The influence of the rotational interaction is investigated
exemplarily by fluid particle dynamics simulations of two trapped particles.
}
\label{sketch_fpd}
\end{figure}

Up to this point we considered finite size effects 
of the particles up to second order  in $a$ (bead radius).
Particles with a finite 
diameter also rotate in shear flows and therefore cause an
additional contribution of third order, which is neglected.
What is more, the rotation is also influenced by
the vicinity of a wall \cite{Brenner:1966.2,Brenner:1966}.
In order to quantify the influence of the rotation
on the wall-induced HI and $\Delta_b$,
we performed computer simulations for finite-sized particles
using the method of fluid particle dynamics (FPD) \cite{Tanaka:2000,Peyla:2007}.
In these simulations the particles are trapped in 
 harmonic potentials  at a distance $d_w=2.5 a$ to
 the wall and, as indicated in Fig.~\ref{sketch_fpd},
 they can freely rotate when exposed to a linear shear flow.
In the FPD simulations for rotating particles
the flow velocity at the particle positions 
 $u_0(d_w)=0.004$ is the same as in  simulations
of the point-like partiles. The squares in Fig.~\ref{delta_dbead_2beads}
represent the results from FPD simulations.

 As displayed in Fig.~\ref{sketch_fpd}
the shear flow profile in  FPD simulations 
was realized by moving the
upper and the lower boundary into opposite 
directions by a constant velocity $v_0$.
For such a configuration the shear induced
particle rotation counteracts the effects of 
a wall (cf. Fig.~\ref{sketch_fpd}). 
In spite of this counteraction  the particle
downstream is again effectively attracted to the wall
and the particle upstream  repelled, similar to 
the case of point-like particles. Accordingly, $\Delta_b$ is
 smaller  for rotating particles than for point like particles of
the same effective radius.
 In Fig.~\ref{2_beads_sketches} and
 Fig.~\ref{sketch_fpd} the shear rates  at the position of the
particles  have an
opposite sign. This has for point like particles no
influence on the elongation  $\Delta_b$. However,
in  FPD simulations of particles of a finite bead-diameter
the bead rotations change their sign with the shear rate.  In contrast
to  the case sketched in Fig.~\ref{sketch_fpd}, 
shear induced bead rotations as indicated in Fig.~\ref{2_beads_sketches}
 support the elongation   $\Delta_b$ 
and $\Delta_b$ becomes in this case larger than for point like particles.

The shear rate in the FPD simulations  was chosen such that the ratio between the
difference of the flow velocity at the lower and at the upper side of the sphere,
$\Delta u_s$,
and the mean velocity $u_0(d_w)$ was $\frac{\Delta u_s}{u_0(d_w)} = \frac{2a \dot\gamma}{u_0(d_w)} = 0.37$.
This causes a reasonably strong particle rotation and therefore a comparatively strong
rotational HI. Despite this quite 
large ratio, the major contribution to $\Delta_b$ is caused 
by the wall-induced HI since $\Delta_b>0$ (cf. Fig.~\ref{delta_dbead_2beads}).
This result supports our approach to use point-like particles
during the rest of this work for the analysis of the 
major trends of the wall-induced HI effects.

$\Delta_b$ becomes large for small values of $d_b$ and $d_w$. In order
to estimate $\Delta_b$ for possible experiments,
we choose $d_b = 3a$ and $d_w = 2a$. Additionally,
a typical potential strength is
$k_{\mathrm{pot}} \simeq 10^{-6} N/m$, the viscosity of water is $\eta \simeq 10^{-3} Ns/m^2$
and typical flow velocities in microfluidic environments are of the order of $u_0(d_w) \simeq 10^{-6} m/s$.
Using these values one obtains
\begin{equation}
\frac{\Delta_b}{a} \simeq \frac{\eta u_0(d_w)}{k_{\mathrm{pot}}} \simeq 10^{-3} \,,
 \label{delta_b_experiment}
\end{equation}
which might be below the currently possible experimental resolution.
However, $\Delta_b$ can be enhanced by using a liquid with a higher viscosity than water
as for example glycerol with $\eta \simeq 1 Ns/m^2$, but then the maximum attainable
flow velocities may be smaller.
Furthermore, the effect can be amplified by placing several beads in a row
and measuring the deflection of the final bead. An amplification of $\Delta_b$ by
about 10\% can be reached by using five beads in a row. Compared to this estimate,
sedimentation experiments close to a wall, as described in section \ref{sec:conclusions}, seem to
be more appropriate for the detection of the 
wall-induced HI effects.

The results discussed until now 
apply to linear shear flow, but the most important property, that the 
streamlines are parallel to the wall, is also shared by other flow profiles
like plane Poiseuille flow.
The spatially varying shear rate in Poiseuille flow causes
higher order effects, but the major results presented here remain qualitatively
valid to Poiseuille flow, as well.

\section{Applications to semiflexible bead-spring chains}\label{sec:appli}
In this section we explore applications of the wall-mediated HI effects
of tethered semiflexible bead-spring models in a flow and fixed near a wall in Sec.~\ref{sec:parallel_chain}
or grafted to the boundary in Sec.~\ref{sec:severalchains}.

\subsection{Polymer model fixed near  a wall}\label{sec:parallel_chain}
\begin{figure}[ht]
  \begin{center}
    \includegraphics[width=0.98\columnwidth]{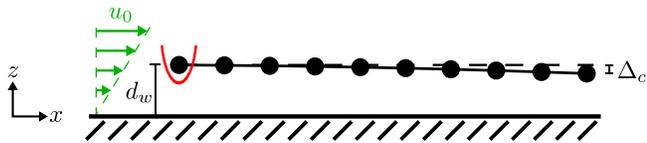}
  \end{center}
  \caption{A semiflexible chain, fixed with one end
at a distance $d_w$ from a wall and exposed to a linear shear flow,
is attracted towards the wall with its free end by a distance
$\Delta_c$. This shift depends on $d_w$ as
shown in Fig.~\ref{delta_dwall_chain_nn}.
For the displayed configuration the parameters are $d_w = 5.5a$
and $u_0(d_w) = 0.004$.
}
\label{chain_bended}
\end{figure}

Polymers tethered with one end in a uniform flow is a model system 
for exploring the importance of HI along polymers at various stages of its 
flow-induced conformations
\cite{Chu:1995.1,Larson:1997.1,Rzehak:1999.2,Rzehak:2000.1,Brochard:1993.1,Kienle:2001.1}.
Here we consider a semiflexible bead-spring polymer model with its
first bead tethered at a distance $d_w$ from a wall as shown in Fig.~\ref{chain_bended}
and we explore the importance of wall-induced hydrodynamic interaction effects.
 
The stationary chain conformation 
in shear flow is again determined via Eq.~(\ref{method_of_reflections})
by using the potential energy for the elastic forces along the polymer
\begin{eqnarray}
V & = & \frac{k_{\mathrm{trap}}}{2} (\vecd{r}_1-\vecd{r}_{\mathrm{trap}})^2 + \sum_{i=1}^{N-1} \frac{k_{\mathrm{str}}}{2} (|\vecd{r}_i-\vecd{r}_{i+1}| - d_{\mathrm{n}})^2 \nonumber \\
&& + \sum_{i=2}^{N-1} \frac{k_{\mathrm{bend}}}{2} \ln{(1+\cos{\chi_i})} \,,
 \label{potential_parallel_chain}
\end{eqnarray}
with   $\vecd{r}_{\mathrm{trap}} = (0,0,d_w)$ the location
of the minimum of the trap potential of strength
$k_{\mathrm{trap}} = 1$.
$k_{\mathrm{str}} = 500$ is the stretching stiffness of the springs
and $d_n=5a$ is the equilibrium distance between neighboring beads.
The bending stiffness is $k_{\mathrm{bend}} = 100$ and
the bending angle at the $i$-th bead is
$\chi_i = \arccos{[ (\vecd{r}_i-\vecd{r}_{i-1}) \cdot (\vecd{r}_{i+1}-\vecd{r}_i) ]}$.
The distances between the beads along the chain are practically fixed on the time scale
of the bending dynamics, which can be seen
by the ratio of the relaxation times of stretching and bending:
$\tau_{\mathrm{str}} / \tau_{\mathrm{bend}} = 2 k_{\mathrm{bend}} / (d_{\mathrm{n}}^2 k_{\mathrm{str}}) \simeq 1/60$.
In the following we consider chains consisting of $N=10$ beads as an example.

In an unbounded fluid the tethered polymer would assume a straight conformation,
parallel to the streamlines of the external flow.
However, in the presence of a wall the free end
of the chain is attracted via the wall-induced HI. This is similar
to the bead attraction downstream investigated in the previous section
and to the hydrodynamically induced  attraction
between two tethered polymers in plug flow and Brownian motion
as  described in Ref.~\cite{Kienle:2011.1}.
A measure of the effective wall  
attraction is the deflection $\Delta_c = d_w - z_N$ 
of the $N$-th bead towards the wall, which is shown in Fig.~\ref{chain_bended} for $d_w = 5.5 a$
and the flow profile (\ref{shearflow_dw}) with $u_0(d_w) = 0.004$.

\begin{figure}[ht]
  \begin{center}
    \includegraphics[width=0.72\columnwidth, angle = -90]{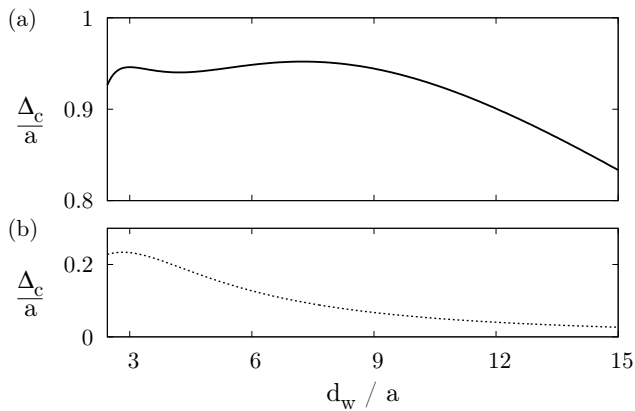}
  \end{center}
  \caption{The deflection $\Delta_c$ of the free end of
a tethered chain consisting of $N=10$ beads (cf. Fig.~\ref{chain_bended})
is shown
as a function of the distance $d_w$ to the wall for the flow field given by
Eq.~(\ref{shearflow_dw}) with $u_0(d_w) = 0.004$.
For the solid line in (a) the wall-mediated
HI between all beads is taken into account.
The dotted line in (b) corresponds to the model assumption 
where the wall-induced HI
has only been taken into account for nearest neighbors and has been
neglected otherwise.
}
\label{delta_dwall_chain_nn}
\end{figure}

$\Delta_c$ is plotted as a function of the wall distance in Fig.~\ref{delta_dwall_chain_nn}
for the constant flow velocity $u_0(d_w)=0.004$ at $z=d_w$.
The solid line in part (a) describes the case
wherein the wall-mediated HI between
all beads is taken into account.
In contrast to the curve for two beads in Fig.~\ref{delta_dwall_2beads},
it displays two maxima and decreases monotonically afterwards.
The first maximum of $\Delta_c(d_w)$ 
at $d_w \simeq 3 a$ is mainly caused 
by the wall-induced HI 
between nearest-neighbor beads along the chain.
In order to substantiate this interpretation,
we plot $\Delta_c (d_w)$ in Fig.~\ref{delta_dwall_chain_nn}(b)
for the model situation where the wall-mediated HI
is only taken into account between 
nearest neighbor beads.
In this case, the magnitude of $\Delta_c(d_w)$ is much
smaller and has indeed only one maximum at $d_w \simeq 2.9a$.
The small shift between this maximum and the first maximum
of the curve in Fig.~\ref{delta_dwall_chain_nn}(a)
is caused by the boundary-induced HI between beads
which are further apart.
The second maximum  at $d_w = 7.3 a$ in Fig.~\ref{delta_dwall_chain_nn}(a) 
is also a result of hydrodynamic bead-bead
interactions via the wall over larger distances than nearest neighbors.

Besides the deflection of the free end,
the chain exhibits also a small curvature.
This can be explained as follows: The trapping potential
prevents the fixed end of the chain from moving away from the wall.
On the other hand each bead is driven to the wall due to
the flow perturbations from all its neighbors upstream.
Therefore, the beads closer to the free
end are increasingly attracted
towards the wall, which leads to a slight bending of the chain.

\subsection{Perpendicularly anchored semiflexible chains in flow}
\label{sec:severalchains}

\begin{figure}[ht]
  \begin{center}
    \includegraphics[width=0.85\columnwidth]{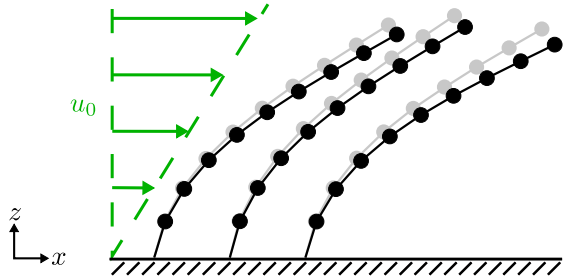}
  \end{center}
  \caption{Three bead-spring chains, which are perpendicularly anchored at a boundary
and exposed to a linear shear flow. For the grey conformation
the  wall-mediated HI is neglected and only the bulk HI effects 
are taken into account.
}
\label{polymer_brush}
\end{figure}

As wall-grafted polymers are relevant in several applications \cite{Stamm:2009}
we investigate the influence of the boundary on the behavior of
semiflexible chains, which are perpendicularly
grafted to a wall and exposed to a shear flow
as shown in Fig.~\ref{polymer_brush}.

For our model calculations we use a potential energy 
describing the bending and stretching of the $N_{\mathrm{c}}$ chains as given by
\begin{eqnarray}
V & = & \sum_{j=1}^{N_{\mathrm{c}}} V^j \quad \mathrm{with} \\
V^j & = & \sum_{i=1}^{N-1} \frac{k_{\mathrm{str}}}{2} (|\vecd{r}_i^j-\vecd{r}_{i+1}^j| - d_{\mathrm{n}})^2 \nonumber \\
&& + \sum_{i=0}^{N-1} \frac{k_{\mathrm{bend}}}{2} \ln{(1+\cos{\chi_i^j})} \,.
 \label{potential_perpendicular_chain}
\end{eqnarray}
Here $\vecd{r}_i^j$ and $\chi_i^j$ are the position vector and the bending angle
of the $i$-th bead in the $j$-th chain.
The constants $k_{\mathrm{str}}$, $d_{\mathrm{n}}$ and 
$k_{\mathrm{bend}}$ have the same values as in the previous subsection,
but the chains are now composed of $N=9$ beads.
However, the second sum in Eq.~\ref{potential_perpendicular_chain} starts with $i=0$ and therefore includes
additional bending contributions at the boundary in order to ensure
that the chain relaxes back to an orientation
perpendicular to the wall after switching off the flow.

A single chain is bent towards the flow direction until the
forces exerted on the beads due to the external drag are
balanced by the stretching and bending forces according to Eq.~(\ref{potential_perpendicular_chain}).
If the flow velocity is very large, the chain even goes beyond
the alignment with the flow lines and bends towards the wall
similar to the results presented in Sec.~\ref{sec:parallel_chain}.

For three chains, which are perpendicularly grafted to a wall, Fig.~\ref{polymer_brush}
shows the steady state chain conformations for the case in which the wall-mediated
contributions to the HI are neglected (grey) and for the case in which the HI via
the boundary is completely included (black).
In the free draining limit, 
where the HI between the beads are disregarded,
the three perpendicularly anchored polymers would be 
bent identically by the flow.
If only the effects of bulk HI
are taken into account,
the outer polymers perturb the flow in
such a way, that the middle polymer experiences
a smaller drag force.
Hence, the first and the last polymer, {\it i.e.} the left and the right one in Fig.~\ref{polymer_brush},
are bent almost identically, while the one in between is bent less strongly,
which is indicated by the grey configuration in Fig.~\ref{polymer_brush}.

Similar to the results from the previous subsection, the wall-mediated HI
leads to a stronger bending of the chains towards the wall, which is displayed by the
deviation of the black chain configurations from the grey ones in Fig.~\ref{polymer_brush}.
This attractive effect increases for chains which lie further downstream,
because the wall-attraction of each chain bead is enhanced by its neighbors
upstream as described in Sec.~\ref{sec:parallel_chain}.
The screening effect for the middle chain due to the interactions in the bulk
is therefore superimposed by the wall-induced effect.

Consequently, the wall-mediated HI along this semi-flexible brush
 leads to a stronger bending of the chains
and thus to a reduction of the brush height.
So the effective diameter of a polymer-grafted tube is increased
and the flow resistance of the grafted chains is reduced,
which might be interesting for studies on wall-grafted brushes of semiflexible
polymers or for a wall decorated by thin flexible pillars.

\section{Conclusions}\label{sec:conclusions} 

In our investigations of wall-induced effects 
on the statics and the dynamics of
hydrodynamically interacting particles
we first calculated the streamline deformation 
around  a single trapped point-like particle close to
 a wall in order to develop a qualitative 
picture about the particle-particle interaction near a wall.
It was shown how the deformations of the flow lines around a fixed particle 
allow an estimation 
of the direction of the force acting on a nearby second particle, which
was confirmed by numerical calculations.

For two or three beads being dragged parallel to a wall,
a scenario with resemblance to sedimenting particles under gravity,
it was shown that the beads always migrate  away from the wall.
The origin of this behavior is due to the interplay between
effects from the bulk and the boundary, and it is similar to the
lift force discussed for vesicles and polymers 
\cite{Misbah:99.1,Seifert:99.1,GrahamM:2004.2,AlexanderK:20011.1}.

The analysis of the configurations of two beads,
which are trapped by harmonic potentials close to a wall and exposed to an external shear flow,
provided further insight about the hydrodynamically mediated particle wall
interaction. We found a repulsion from the wall for the particle upstream
and an attraction towards the boundary for the one downstream.
Varying the particle-wall distance and the particle-particle distance a
characteristic maximum in the deflection of the downstream particle  was found,
which could be  described also analytically giving further insight
on the  parameter dependence of this phenomenon. 
 The behavior obtained for two point-like particles 
was confirmed  by using 
the complementary method of Fluid Particle Dynamics 
 \cite{Tanaka:2000,Peyla:2007},
which accounts for the finite particle radii and the effects
of particle rotation on the hydrodynamic particle-particle interaction.

As an application of the basic
effects found for the two-bead configuration,
 we investigated a semiflexible bead-spring chain, 
where one end is held in a linear shear flow at a
distance $d_w$ from the boundary.
The chain is attracted towards the boundary as a function of
$d_w$ and we
identified the wall-induced contributions to the HI as the source
for this behavior. The phenomenon is related to a recent study on
the flow induced polymer-polymer attraction \cite{Kienle:2011.1}
mediated through inter-chain HI,
where the second polymer causes very similar effects as 
the boundary in this  work.  Both are
examples of hydrodynamically induced particle-particle attraction,
which has recently  been found for rotors as well \cite{Schreiber:2010.1}.

What is more, three perpendicularly anchored semiflexible bead-spring chains
were investigated as a simple model for a polymer brush 
and their response to an external linear shear flow was obtained. 
It was found , that  wall-effects cause a stronger effective attraction
towards the wall for the polymers downstream than for   the ones upstream. 
Whether three semiflexible polymers, perpendicularly anchored at a wall,
but not along a line, 
and of different length, have also the propensity to  oscillatory motion as 
reported for three
beads in shear flow in Ref.~\cite{Holzer:2006.1}, is an interesting further question.

While our analysis in this work is exclusively used
for a linear shear flow  as external stimulus,
we nonetheless expect a similar qualitative behavior for
other flow profiles with parallel streamlines. 
The reason is, that the deformations of the streamlines
near walls  as shown in Fig.~\ref{streamlines} 
are in general the same for other laminar flow profiles
with parallel streamlines.

Our results may be tested by different experiments.
A first one would be to measure the wall-induced 
displacement of an array of beads trapped close to wall
by laser tweezers  while imposing
 an external flow similar to the 
setups shown in Fig.~\ref{2_beads_sketches} and Fig.~\ref{chain_bended}. 
A variation of the above setup may be to measure
the deflection of a cantilever in proximity of a wall and exposed to
a flow as illustrated for a model polymer in Sec.~\ref{sec:parallel_chain}.

Another alternative  to probe  our findings is to line up
two particles in a row close to the wall of a container, extended
in its vertical direction,  and to track the 
trajectories of the two sedimenting particles.
According to our predictions,  the  particle moving in front should be
repelled from the wall whereas the particle behind should at first be attracted.
As soon as the connection vector between the particles becomes sufficiently oblique
with respect to the boundary, 
the bulk effect is expected to become 
dominant, so that both beads
are carried both away from the wall.

\acknowledgments{Inspiring discussions with Diego Kienle and  Chaouqi  Misbah are  
acknowledged. This work has been supported by the German science foundation (through
FOR 608 and SPP 1164) and by the Bayerisch-Franz\"osisches Hochschulzentrum (BFHZ).
}


\vspace{6mm}

\begin{thebibliography}{10}
\bibitem{Brenner:1991}
J.~Happel and H.~Brenner, 
{\it Low Reynolds Number Hydrodynamics} (Prentice-Hall, Englewood Cliffs, 1981).

\bibitem{Russel:1989}
W.~B. Russel, D.~A. Saville, and W.~R. Showalter,    {\it Colloidal Dispersions}
(Cambridge University Press, Cambridge, 1989).

\bibitem{Dhont:96} J.~K.~G. Dhont, {\it An Introduction to dynamics of colloids}
(Elsevier, Amsterdam, 1996).

\bibitem{Leal:2007} L.~G. Leal, {\it Advanced Transport Phenomena},
(Cambridge University Press, Cambridge, 2007).



\bibitem{Silberberg:61.1}
G. Segr\'e and A. Silberberg, ''Radial Poiseuille flow of suspensions,'' Nature {\bf 189}, 209 (1961).

\bibitem{Silberberg:62.1}
G. Segr\'e and A. Silberberg, ''Behavior of macroscopic rigid spheres in
                           Poiseuille flow. Part 1. Determination of local concentration
                           by statistical analysis of particle passages through crossed light 
                           beams,''  J. Fluid Mech. {\bf 14}, 115 (1962).


\bibitem{Leal:1980.1}
L.~G. Leal, ''Particle motions in a viscous fluid,'' Ann. Rev. Fluid Mech. {\bf 12}, 435 (1980).

\bibitem{Hinch:1989.1}
J.~A. Schonberg and E.~J. Hinch,
''Inertial migration of solid particles in Poiseuille flow - I. Theory,''
J. Fluid Mech. {\bf 203}, 517 (1989).



\bibitem{JosephDD:2005.1}
B.~H. Yang, J. Wang, D.~D. Joseph, H.~H. Hu, T.~W. Pan, and R. Glowinski, 
''Migration of a sphere in tube flow,''
J. Fluid Mech. {\bf 540}, 109 (2005).

\bibitem{Stone:2009.1} D. Di~Carlo, J.~F. Edd,
K.~J. Humphry, H.~A. Stone, and M. Toner,
''Particle segregation and dynamics of confined flows,''
Phys. Rev. Lett. {\bf 102},
094503 (2009).



\bibitem{Seifert:99.1}
U. Seifert, ''Hydrodynamic lift on bound vesicles,'' Phys. Rev. Lett. {\bf 83}, 876 (1999).



\bibitem{Misbah:99.1}
I. Cantat and C. Misbah, ''Lift force and dynamically unbinding of adhering
vesicles under shear flow,'' Phys. Rev. Lett. {\bf 83}, 880 (1999).

\bibitem{Viallat:2002.1}
 M. Abkarian, C. Lartigue, and A. Viallat, ''Tank Treading and unbinding of deformable
vesicles in shear flow: Determination of the lift force,'' Phys. Rev. Lett. {\bf 88}, 068102 (2002).

\bibitem{Kaoui:2008.1}
B. Kaoui, G.~H. Ristow, I. Cantat, C. Misbah, and W. Zimmermann,
''Lateral migration of a two-dimensional vesicle in unbounded
Poiseuille flow,'' Phys. Rev. E {\bf 77}, 021903 (2008).

\bibitem{Kaoui:2008.2}
G. Coupier, B. Kaoui, T. Podgorski, and C. Misbah,
''Noninertial lateral migration of vesicles in bounded
Poiseuille flow,'' Phys. Fluids {\bf 20}, 111702 (2008).

\bibitem{Agarwal:1994.1}
U.~S. Agarwal, A. Dutta, and R.~A. Mashelkar,
''Migration of macromolecules under flow: 
                           The physical origin and engineering implications,''
Chem. Eng. Sci. {\bf 49}, 1693 (1994).



\bibitem{GrahamM:2004.2}
R.~M. Jendrejack,  D.~C. Schwartz,  J.~J. de~Pablo, and M.~D. Graham,
''Shear-induced migration in flowing polymer solutions: Simulation
                    of long-chain DNA in microchannels,''
J. Chem. Phys. {\bf 120}, 2513 (2004).


\bibitem{GrahamM:2006.1}
J.~P Hern\'andez-Ortiz and H. Ma and J.~J. de~Pablo, and M.~D. Graham,
''Cross-streamline migration in confined flowing polymer solutions:
                    Theory and simulation,''
Phys. Fluids {\bf 18}, 123101 (2006).


\bibitem{GrahamM:2011.1} M. Graham, ''Fluid dynamics of dissolved
polymer molecules in confined geometries,'' 
Annu. Rev. Fluid Mech. {\bf 43}, 273 (2011).


\bibitem{Chelakkot:2010.1}
R.~Chelakkot, R.~G.~Winkler, and G.~Gompper, ''Migration of semiflexible
polymers in microcapillary flow,'' EPL {\bf 91}, 14001 (2010).


\bibitem{Lundgren:1987}
A.~F. Fortes, D.~D. Joseph, and T.~S. Lundgren, ''Nonlinear mechanics
of fluidization of beds of spherical particles,'' J. Fluid Mech. {\bf 177}, 467 (1987).

\bibitem{Nissila:2004.1}
E. Kuusela, J.~M. Lahtinen, and T. Ala-Nissila, ''Sedimentation dynamics of spherical
particles in confined geometries,'' Phys. Rev. E {\bf 69}, 066310 (2004).




\bibitem{Powers:2009}
E. Lauga and T.~R. Powers, ''The hydrodynamics of swimming microorganisms,'' 
Rep. Prog. Phys. {\bf 72}, 096601 (2009).

\bibitem{Lauga:2008}
A.~P. Berke, L. Turner, H.~C. Berg, and E.~Lauga,
''Hydrodynamic attraction of swimming microorganisms by surfaces,''
 Phys. Rev. Lett. {\bf 101}, 038102 (2008).




\bibitem{Minko:2006}
S.~Minko, ''Responsive polymer brushes,''
 J. Macromol. Sci. C: Pol. Rev., {\bf 46}, 397 (2006).

\bibitem{Brenner:2005}
S.~P. Adiga and  D.~W. Brenner,
''Flow control through polymer-grafted smart nanofluidic channels: Molecular dynamics simulations,'' 
 Nanolett. {\bf 5}, 2509 (2005).


\bibitem{Stamm:2009}
P. Uhlmann, H. Merlitz, J.~U. Sommer, and  M. Stamm, 
''Polymer brushes for surface tuning,'' Macromol. Rapid Comm. {\bf 30}, 732 (2009).


\bibitem{Viovy:2000}
P.~S. Doyle, B. Ladoux, and J.-L. Viovy,
''Dynamics of a tethered polymer in shear flow,''
 Phys. Rev. Lett. {\bf 84}, 4769 (2000).

\bibitem{Chu:2005}
C.~M. Schroeder, R.~E. Teixeira, E.~S.~G. Shaqfeh, and S. Chu,
''Characteristic periodic motion of polymers in shear flow,''
 Phys. Rev. Lett. {\bf 95}, 018301 (2005).


\bibitem{Delgado:2006.1}
R. Delgado-Buscalioni, ''Cyclic motion of a grafted polymer under
                    shear flow,'' Phys. Rev. Lett.
{\bf 96}, 088303 (2006).
 



\bibitem{Shaqfeh:2009}
C.~A. Lueth and  E.~S.~G. Shaqfeh, 
''Experimental and numerical studies of tethered DNA shear dynamics 
                         in flow-gradient plane,''
Macromolecules {\bf 42}, 9170 (2009).



\bibitem{dePablo:2009}
Y. Zhang, A. Donev, T. Weisgraber, B.~J. Alder, M.~D. Graham, and J.~J. de Pablo, 
''Tethered DNA dynamics in shear flow,''
J. Chem. Phys. {\bf 130}, 234902 (2009).


\bibitem{Loewen:2009}
G.~L. He, R.~Messina, H.~L\"owen,
 A.~Kiriy, V.~Bocharova, and M.~Stamm, 
''Shear-induced stretching of adsorbed polymer chains,''
Soft Matter {\bf 5}, 3014 (2009).

\bibitem{AlexanderK:20011.1}
C.~E. Sing and A. Alexander-Katz,
''Non-monotonic hydrodynamic lift force on highly extended
polymers near surfaces,''
EPL {\bf 95}, 48001 (2011).


\bibitem{Holzer:2006.1} L. Holzer and W. Zimmermann, 
''Particles held by springs in a linear shear flow exhibit oscillatory
motion,'' Phys. Rev. E {\bf 73}, 060801 (R) (2006).
 

\bibitem{Carlsson_A:2007}
A. Carlsson, F. Lundell, and L. S\"oderberg, ''Fiber orientation control
related to papermaking,'' J. Fluids Eng. {\bf 129}, 457 (2007).

\bibitem{Caflisch:1988.1} R.~E. Caflisch, C. Lim, J.~H.~C. Luke,
and A.~S. Sangami, ''Periodic solutions for three sedimenting
spheres ,'' Phys. Fluids {\bf 31}, 3175 (1988).



\bibitem{Schreiber:2010.1}
S. Schreiber,  T. Fischer, and W. Zimmermann,
''Hydrodynamic attraction and repulsion between 
                             asymmetric rotors,''
New J. Phys. {\bf 12}, 073017 (2010).


\bibitem{Kienle:2011.1} D. Kienle, R. Rzehak, and W. Zimmermann,
''Identifying hydrodynamic interaction effects in tethered 
polymers in uniform flow,''
Phys.  Rev. E {\bf 83}, 062802 (2011).



\bibitem{Zarraga:2002.1}
I.~E. Zarraga and D.~T. Leighton,
''Measurement of an unexpectedly large shear-induced
                         self-diffusivity in a dilute suspension of spheres,''
Phys. Fluids {\bf 14}, 2194 (2002).

\bibitem{Wajnryb:2007.1}
M. Zurita-Gotor, J. Blawzdziewicz, and E. Wajnryb,
''Swapping trajectories: a new wall-induced cross-streamline
particle migration mechanism in a dilute suspension of spheres,''
J. Fluid Mech. {\bf 592}, 447 (2007).





\bibitem{Blake:1971}
J.~R. Blake,
''A note on the image system for a stokeslet in a no-slip boundary,''
 Proc. Camb. Phil. Soc. {\bf 70}, 303 (1971).

\bibitem{Wajnryb:2000.1}
B. Cichocki, R.~B. Jones, R. Kutteh, and E. Wajnryb,
''Friction and mobility for colloidal spheres
                           in Stokes flow near a boundary:
                           The multipole method and applications,''
J. Chem. Phys. {\bf 117}, 2548 (2000).



\bibitem{Brady:2007}
J.~W. Swan and J.~F. Brady, ''Simulation of hydrodynamically interacting particles near a no-slip boundary,''  Phys. Fluids {\bf 19}, 113306 (2007).

\bibitem{Tanaka:2000}
H.~Tanaka and T.~Araki, ''Simulation method of colloidal suspensions with 
                           hydrodynamic interactions: Fluid particle dynamics,''
Phys. Rev. Lett. {\bf 85}, 1338 (2000).


\bibitem{Peyla:2007}
P.~Peyla, 
''Rheology and dynamics of a deformable object in a microfluidic configuration: a numerical study,''
EPL {\bf 80}, 34001 (2007); Y. Davit and P. Peyla, ''Intriguing viscosity effects in confined suspensions
                          : a numerical study,'' EPL {\bf 83}, 64001 (2008);
Jibuti, S. Rafai and P. Peyla, ''Suspensions with a tunable effective viscosity:
                          a numerical study,'' J. Fluid Mech. {\bf 693}, 345 (2012).

\bibitem{canti}
T. Naik, E. Longmire, and S. Mantell,
''Dynamic response of a cantilever in liquid near a solid wall,'' 
 Sensors and Actuators, {\bf 102}, 240 (2007).



\bibitem{Brenner:66} 
A.~J. Goldman, R.~G. Cox, and H. Brenner,
''The slow motion of two identical arbitrarily 
                            oriented spheres through a viscous fluid,''
 Chem. Eng. Sci. {\bf 21}, 1151 (1966).

\bibitem{Brenner:1966}
A.~J.~Goldman, R.~G.~Cox, and H.~Brenner, 
''Slow viscous motion of a sphere parallel to a plane wall – I Motion through a quiescent fluid,''
Chem. Eng. Sci {\bf 22}, 637 (1967).


\bibitem{Brenner:1966.2}
A.~J. Goldman, R.~G. Cox, and H. Brenner,
''Slow viscous motion of a sphere parallel to a plane wall – II Couette flow,''
 Chem. Eng. Sci {\bf 22}, 653 (1967).





\bibitem{Chu:1995.1} T.~T. Perkins, D.~E. Smith,  R. Larson, and S. Chu,
 ''Stretching of a Single Tethered Polymer in 
                           a Uniform Flow,''
 Science {\bf 268}, 83 (1995).


\bibitem{Larson:1997.1} R.~G. Larson, T.~T. Perkins, D.~E. Smith, and S. Chu,
   ''Hydrodynamics of a {DNA} molecule in a flow field,'' 
Phys. Rev. E {\bf 55}, 1794 (1997).


\bibitem{Rzehak:1999.2} R. Rzehak,  D. Kienle,  T. Kawakatsu, and W. Zimmermann,
''Partial draining effect of a tethered polymer in flow,''
 Europhys. Lett. {\bf 46}, 821 (1999).

\bibitem{Rzehak:2000.1} R. Rzehak,  W. Kromen,  T. Kawakatsu, and W. Zimmermann,
  ''Deformations of tethered polymers in uniform flow,'' 
  Eur. Phys. J. E. {\bf 2}, 3 (2000).


\bibitem{Brochard:1993.1} F. Brochard-Wyart, 
 ''{D}eformations of one tethered chain in strong flows,''
Europhys. Lett. {\bf 23}, 105 (1993).

\bibitem{Kienle:2001.1} D. Kienle and W. Zimmermann,
  ''F-Shell Blob-Modell for a Tethered
                            Polymer in Strong Flows,''
Macromolecules {\bf 34}, 9173 (2001).





\end{thebibliography}
\end{document}